\documentclass[12pt,preprint]{aastex}
\newcommand{\myemail}{lychiang@asiaa.sinica.edu.tw}
\newcommand{\bi}[1]{\mbox{\boldmath $#1$}}
\usepackage{float}
\usepackage{amsmath}
\usepackage{epsfig}
\usepackage{subfigure}
\makeatletter

\newcommand{\beq}{\begin{equation}}
\newcommand{\eeq}{\end{equation}}
\newcommand{\be}{\begin{eqnarray}}
\newcommand{\ee}{\end{eqnarray}}

\def\uks{{\mu{\rm K}^2}}

\def\wmap{{\sl WMAP }}

\def\healpix{H{\sc ealpix }}
\def\glesp{G{\sc lesp }}
\def\l{{\ell}}

\def\lm{{\l m}}
\def\cl{{C_\l}}
\def\xl{{X_\l}}
\def\lcdm{{\Lambda}{\rm CDM}}
\def\dtq{{\delta T_\l^2}}
\slugcomment{Submitted to Astrophysical Journal}

\shorttitle{Cross-Power Spectrum and Its Application}
\shortauthors{Fei-Fan Chen and Lung-Yih Chiang}

\begin{document}
\title{Cross-Power Spectrum and Its Application on Window Functions in the WMAP Data}
\author{Lung-Yih Chiang}
\affil{\asiaa}
\author{Fei-Fan Chen}
\affil{\ntu}
\affil{\asiaa}

\email{\myemail}
\newcommand{\ntu}{{Institute of Astrophysics, National Taiwan University, 1, Rooservolt Road, Taipei, Taiwan}}
\newcommand{\asiaa}{{Institute of Astronomy and Astrophysics, Academia Sinica, P.O. Box 23-141, Taipei 10617, Taiwan}}

\begin{abstract}
Cross-power spectrum is a quadratic estimator between two maps that can provide unbiased estimate of the  underlying power spectrum of the correlated signals, which is therefore used for extracting the power spectrum in the WMAP data. In this paper we discuss the limit of cross-power spectrum and derive the residual from uncorrelated signal, which is the source of error in power spectrum extraction. We employ the estimator to extract window functions by crossing pairs of extragalactic point sources. We desmonstrate its usefulness in WMAP Difference Assembly maps where the window functions are measured via Jupiter and then extract the window functions of the 5 WMAP frequency band maps.

\end{abstract}

\section{Introduction}
Variables and parameters in theories of cosmology are often derived based on a statistical ensemble. In practice, particularly in data analysis on the CMB, however, one often encounters the well known and rather unique issue, namely we can do statistical analysis on only one version of the celestial sphere using an ergodic hypothesis. For example, the angular power spectrum of the temperature anisotropies $\cl\equiv \langle a^{}_\lm a^{*}_\lm\rangle$, but with one version of the sphere it can be estimated with only $2\l+1$ modes: $\cl^{\rm est}=(2\l+1)^{-1} \sum_{m=-\l}^{\l} a^{}_\lm a^{*}_\lm$.
Lack of a statistical ensemble thus induces in power spectrum analysis the  uncertainty called ``cosmic  variance'' \citep{cv}. Another issue arising from lack of an ensemble is coined ``cosmic covariance'', which refers to a non-zero chance correlation between the CMB and the foregrounds \citep{cc}. The cosmic covariance is the source of error in employing quadratic minimization for separating the CMB and the foregrounds \citep{wmap1yfg,wmap3ytem}. 

In this paper we address another issue in cross-power spectrum arising from lack of an ensemble. Cross-power spectrum is useful in eliminating uncorrelated signal and is thus commonly used in power spectrum estimation from multifrequency measurement of CMB. It is required, however, to have an ensemble of modes in the summation to completely eliminate the uncorrelated signal. Therefore lack of an ensemble of modes results in residual (hence error) in the cross-power spectrum. 

Another important issues related to extraction of the power spectrum of CMB temperature anisotrpies is the calibration of window function convolving the signal, which is usually by measuring Jupiter \citep{wmap1ybeam}. It can also be via measuring bright extragalactic point sources as they are manifestation of the window function as well, albeit with background CMB and instrument noise. We can then employ cross-power spectrum on pairs of extragalactic point sources from different patches of the sky, where the background CMB and other signals are uncorrelated and can be largely eliminated.  

This paper is arranged as follows. In Section 2 we examine the limit of cross-power spectrum as a quadratic estimator. We in Section 3 employ the estimator on a pair of simulated point sources to obtain the window function. In Section 4 we put this method to the test on WMAP Difference Assembly (DA) maps whose window functions are already carefully calibrated via measuring Jupiter and we extract the window functions of WMAP frequency band maps. The conclusion and discussion is in Section 5.

\section{Cross-power spectrum and its limit}
Cross-power spectrum (XPS) is a quadratic estimator between two maps $j$ and $j'$, and on a sphere it is written as
\begin{equation}
\xl^{jj'}=\frac{1}{2\l+1} \sum^\l_{m=-\l} j^{*}_\lm j'_\lm,
\label{def}
\end{equation}
where $j_\lm$ and $j'_\lm$ are the spherical harmonic coefficients of the two maps, and $*$ denotes complex conjugate \citep{wmap1ypower}. In order for the XPS to be real, the usual practice is by replacing the $j_\lm^* j'_\lm$ in Eq.(\ref{def}) with
\begin{equation}
( j^{*}_\lm j'_\lm+j'^{*}_\lm j_\lm)/2\equiv |j_\lm||j'_\lm|\cos \Delta \phi_\lm,
\label{prac}
\end{equation}
where $\Delta \phi_\lm$ is the phase difference between $j_\lm$ and $j'_\lm$. The advantage of XPS as an unbiased quadratic estimator for power spectrum estimation lies in the fact that XPS returns with its usual power spectrum $\sum_m |j_\lm|^2$ if $j$ and $j'$ are of the same signal, and if, on the other hand, $j$ and $j'$ are uncorrelated then $\Delta \phi_\lm$ is distributed uniformly random in $[0, 2\pi]$ and thus $\langle\sum_{m=-\l}^{\l} |j_\lm||j'_\lm|\cos \Delta \phi_\lm \rangle=0$, where the angle brackets denote ensemble average. Hence XPS is useful in eliminating uncorrelated signals while preserving the correlated one and has been employed by WMAP to extract CMB spectrum by crossing the Differencing Assemblies (DA) \citep{wmap1ypower,wmap3ytem,wmap5ypower,wmap7ypower} . In practice, however, one does not have a statistical ensemble and the capability of XPS to extract the underlying  power spectrum from correlated signal depends on how much the uncorrelated signal is decreased with finite realizations and finite number of multipole modes involved in the summation at $\l$.

Take as an example extracting the CMB signal from \wmap data. The W1 and W2 DA have the same CMB but uncorrelated noise, so writing $a_\lm^{\rm W1}=a_\lm^{\rm c}+ n_\lm$ and $a_\lm^{\rm W2}=a_\lm^{\rm c}+ n'_\lm$, in XPS the correlated signal $\sum |a^{\rm c}_\lm|^2$ is the desired one whereas those uncorrelated terms between CMB and noises $\xl^{\rm cn}$, $\xl^{\rm cn'}$ and between noises $\xl^{\rm nn'}$ shall be decreased but nevertheless not to zero,  which is then the source of error in the XPS.

\begin{figure}
\plotone{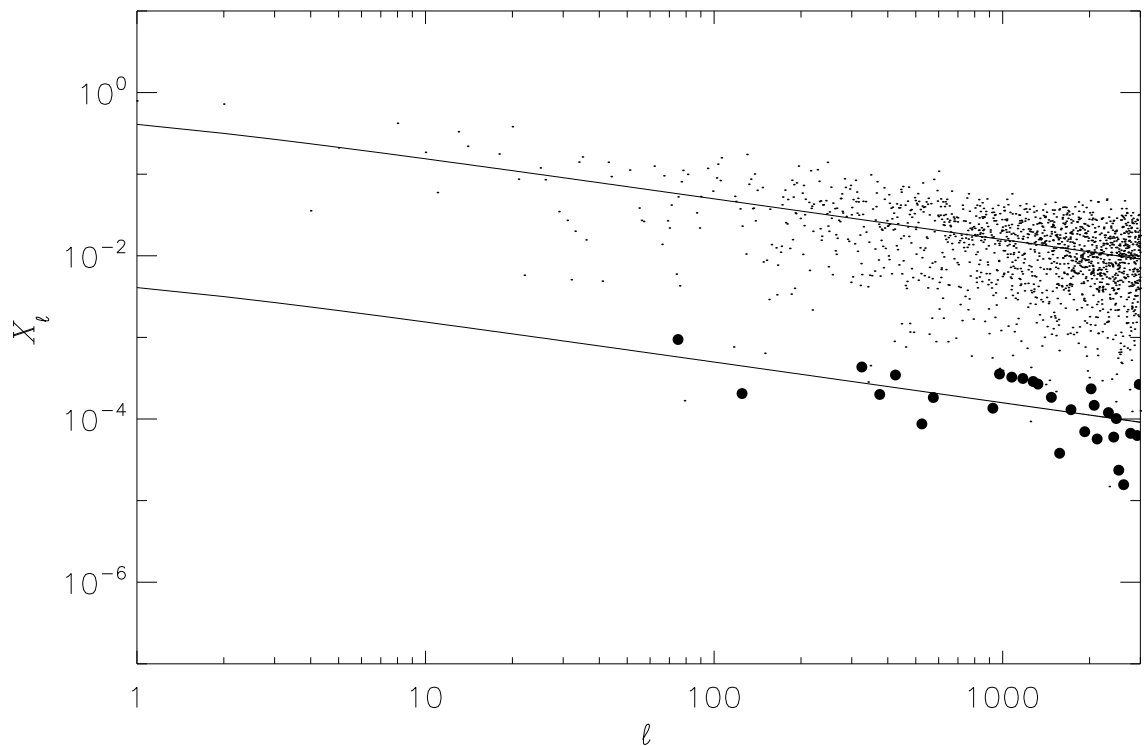}
\caption{Demonstration of Eq.(\ref{result}). The black dots denote XPS from two white noise maps with unity power spectrum, whereas the larger dots are the average from 200 pairs of white noise maps, binned with $ L \equiv \Delta \l=50$. The upper line is $(2\sqrt{\l+1/2})^{-1}$ and the lower $(2\sqrt{LN(\l+1/2)})^{-1}$, where $L=50$ and $N=200$. Note that the negative values are not plotted. }
\label{demo}
\end{figure}

In order to examine the level of residual in XPS, we consider two Gaussian and uncorrelated maps $j$ and $j'$, whose spherical harmonic coefficient $j_\lm$ and $j'_\lm$ have zero mean and variance $J_\l$ and $J'_\l$ respectively and the corresponding phase difference $\Delta \phi_\lm$ is uniformly random in $[0, 2\pi]$. One should note that the distribution of Eq.(\ref{prac}) under this condition is a normal product distribution, which is itself not Gaussian but can be expressed as 
\begin{equation}
p_{j _\lm j'_\lm}(x)=\frac{4 K_0\left(\left| x \right|/\sqrt{J _\l J'_\l}\right)}{\pi\sqrt{J_\l J'_\l}},
\end{equation}
where $K_0$ is the zeroth-ordered modified Bessel function of the 2nd kind \citep{springer}. Although the distribution of the variable is not Gaussian, what we need to examine is its summation. Since the XPS defined in Eq.(\ref{prac}) can be either positive or negative, we look instead at its root mean square $\sqrt{ \langle (\xl^{jj'})^2 \rangle}$. Due to equal partiution on both perpendicular directions of complex plane, it can be written:
\begin{equation}
\sqrt{ \langle (\xl^{jj'})^2 \rangle}=\frac{1}{2\l+1}\sqrt{\frac{1}{2}\Bigg\langle\left| \sum_{m=-\l}^\l |j_\lm||j'_\lm| \exp(i \Delta \phi_\lm) \right|^2 \Bigg\rangle},
\label{randomwalk}
\end{equation} 
the right hand side of which is closely related to a two-dimensional random walk. Random walk statistics is employed as a Gaussianity test on CMB \citep{naselsky,coles}. Recall that the celebrated Pearson's random walk is a 2D isotropic random walk with {\it equal} step-length, expressed in the complex plane ${\bi r}=\sum_k |{\bi d}_k| \exp(i\phi_k)$ where $|{\bi d}_k|=d$. The pdf of the resultant displacement $r\equiv |{\bi r}|$ after $\l$ steps has the form 
\begin{equation}
p(r)=\frac{1}{\pi \l d^2} \exp\left(-\frac{r^2}{\l d^2}\right),
\label{pearson}
\end{equation}
from which $\langle r^2\rangle=\l d^2$. Pearson's walk can be generalized to unequal step-length and it can  be shown \citep{pearson,rayleigh} that when $\l$ is large and the mean-square step length is known : $(\l)^{-1}\sum^\l |{\bi d}_k|^2=d^2$, the pdf for the displacement has the same form as Eq.(\ref{pearson}) \citep{hughes}. For unequal step length $|{\bi d}_k|=|j_\lm||j'_\lm|$, the ensembled-average square step length is then $ \langle |j_\lm|^2|j'_\lm|^2 \rangle =J_\l J'_\l$, so the mean square displacement of the random walk $\langle| \sum_{m=-\l}^\l |j_\lm||j'_\lm| \exp(i \Delta \phi_\lm)|^2\rangle = (2\l+1)J_\l J'_\l $. Note that in driving the result we only assume $j$ and $j'$ are Gaussian, so they can be either white noise (with flat power spectrum) or others whose power spectrum has scale dependence. In any case, we can write
\begin{equation}
\frac{\sqrt{\langle (\xl^{jj'})^2 \rangle} }{\sqrt{ J_\l J'_\l }}= \frac{1}{2\sqrt{\l+1/2}}.
\label{result}
\end{equation}
The decreasing of the ``junk'' is inversely proportional to the square root of the number of random walk, and it can be further decreased by $1/\sqrt{LN}$ with binning $L\equiv \Delta \l$ multipole numbers and averaging from $N$ sets of XPS. We demonstrate Eq.(\ref{result}) in Fig.\ref{demo} by plotting $X_\l$ from two white-noise maps. One can see the the decreasing of the power spectrum proportional to $(2\sqrt{\l+1/2})^{-1}$. Note that $X_\l$ is jumping either postively or negatively and only positive points are plotted. 

Knowing the residual of the junk from XPS, we can estimate its capability in extracting CMB. In crossing $N$ pairs of maps containing CMB with the same noise level, we assume the CMB power spectrum is $C_\l \simeq A\l^{-2}$ and the instrument noise level is $N_\l=RA$ where $R$ indicates the noise level relative to the CMB and $R\ll 1$\footnote{The noise level $N_\l=RA$ is related to noise fluctuation $\sigma^2$ through $\sigma^2=N_{\rm pix} RA/ 4\pi$, where $N_{\rm pix}$ is the total pixel number of the map.}. The residual level from between CMB and noise, and between noises are $2\xl^{\rm cn}\simeq A \sqrt{R}(\l^3LN)^{-1/2} $ and $\xl^{\rm nn'} \simeq RA (4\l LN)^{-1/2} $, respectively. They intersect at $\l \simeq 2/\sqrt{R}$. For $\l <2/\sqrt{R}$, $\xl^{\rm cn}$ is the main residual with contribution: $2\xl^{\rm cn}/\cl^{\rm c}\simeq \sqrt{R\l/LN}$, and $LN>\l^2 R$ in order for the residual contribution to be within cosmic variance, which shall be the case as long as $LN \ge 4$. For $\l \ge 2/\sqrt{R}$, on the other hand, the residual in the CMB extraction is $R \sqrt{\l^3/4LN}$, and beyond $\l=(4LN/R^2)^{1/3}$ the residual level is higher than the CMB, thereby XPS fails.

\begin{figure}
\centering
\plotone{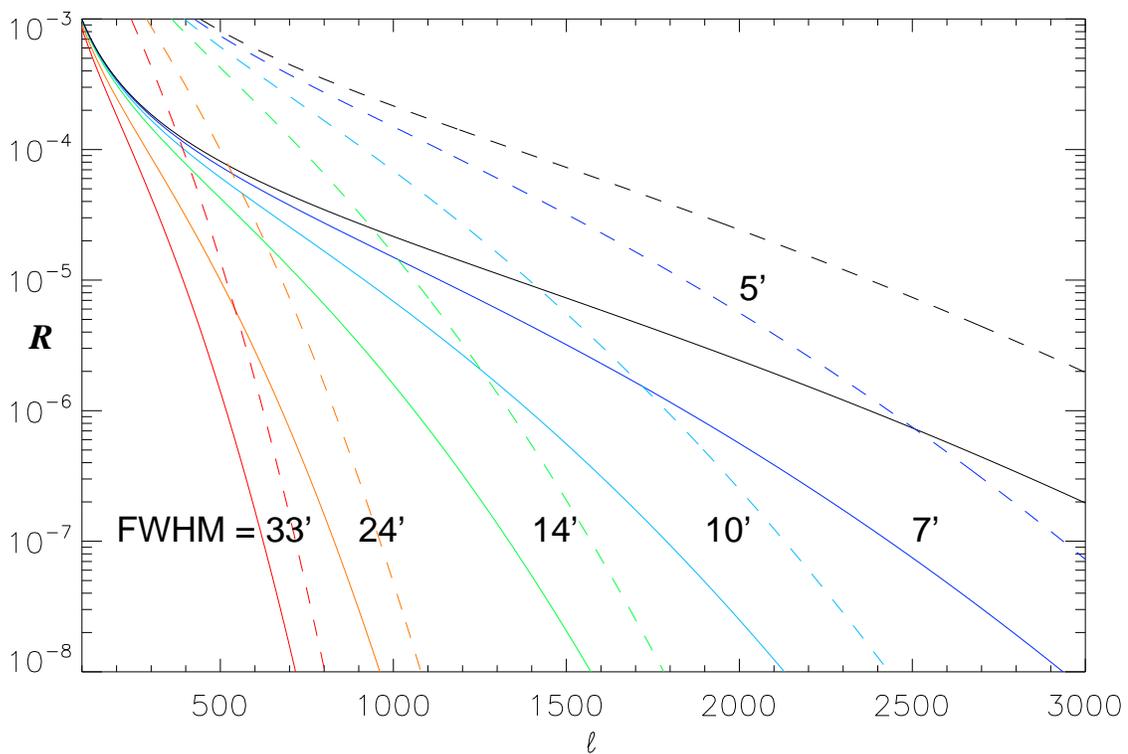}
\caption{The upper limit of $R$, which corresponds to the instrument noise level, for the XPS to be able to successfully retrieve the CMB at the corresponding multipole number. Red, orange, green, light blue, blue and black color denote FWHM$=33, 24, 14, 10, 7, 5$ arcmin, respectively, which then correspond to ESA Planck Surveyor LFI 30, 44, 70 GHz channel, HFI 100, 143 GHz, and the rest channels with higher frequency. The solid and dash lines are for $LN=1$ and 100, respectively. }
\label{beamlimit} 
\end{figure}

For a more realistic situation, where the XPS is employed on signals that is convolved with window function: $A \l^{-2} \exp(-B^2 \l^2)$, we plot in Fig.\ref{beamlimit} the upper limit of $R$ (the noise level) if the XPS is to retrieve the CMB at the corresponding multipole number. The different FWHM are chosen according to those from channels of ESA Planck Surveyor. As expected, more binning and more sets of XPS  shall retrieve higher modes for the same noise level.


\begin{figure}
\plottwo{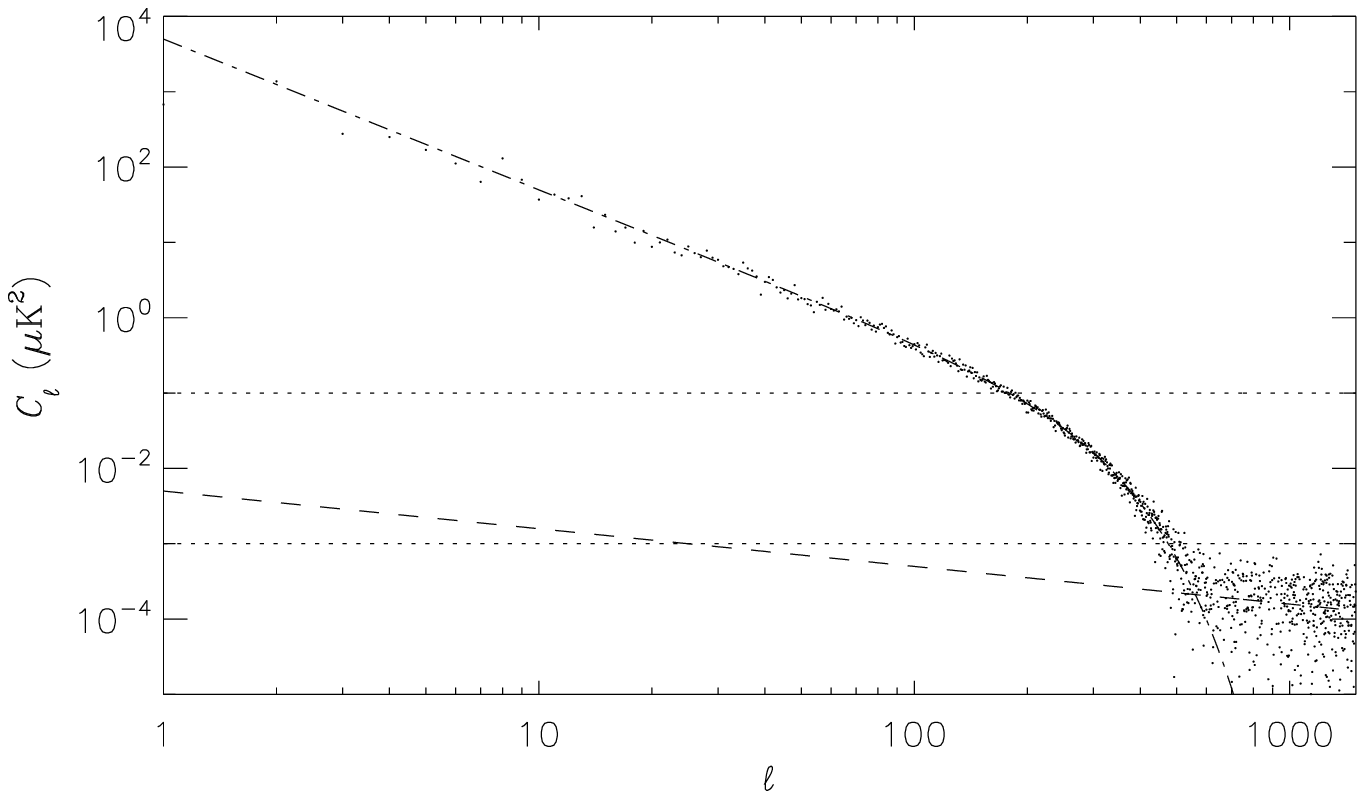}{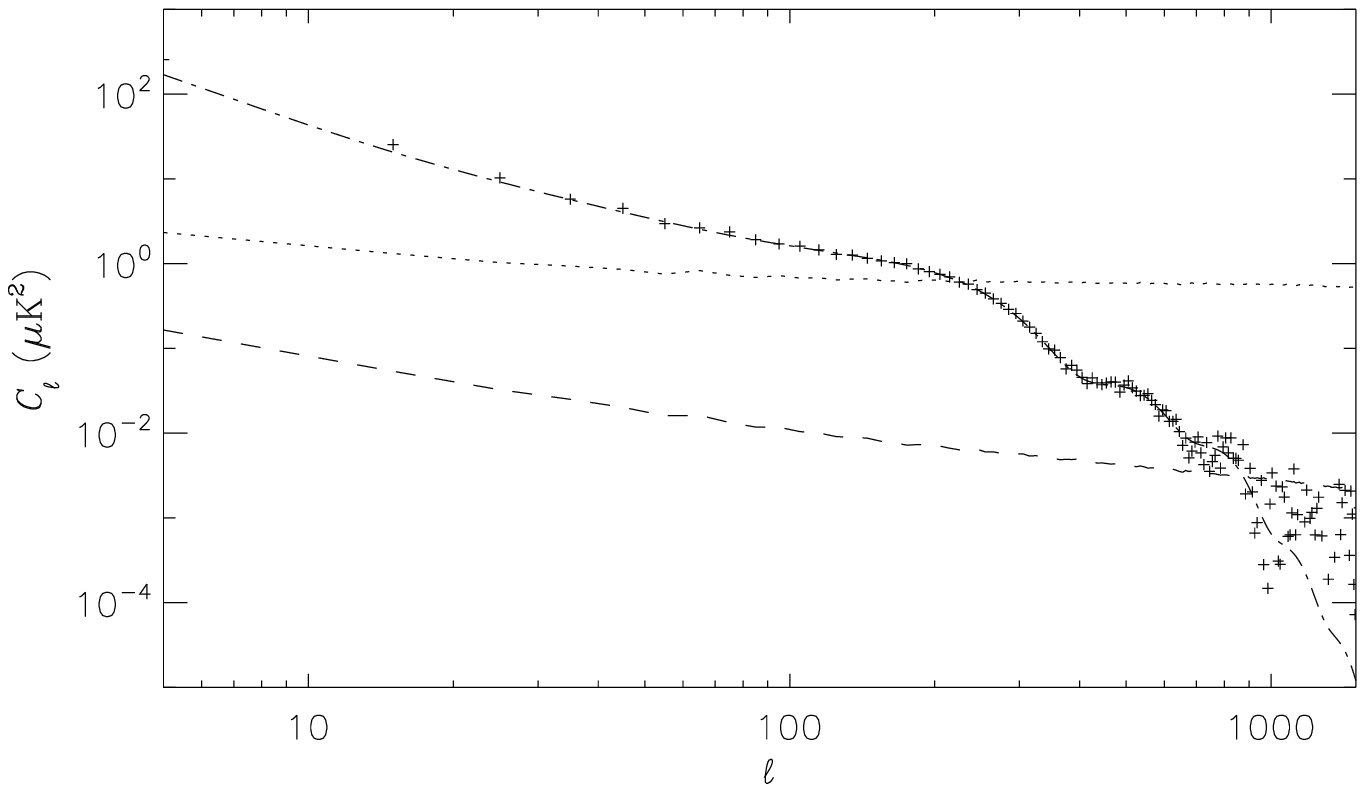}
\caption{Demonstration of the XPS retriving power and its limit. Top panel shows XPS (dots) of two maps with input spectrum $\cl=10^{4} \l^{-2} (\uks)$ convolved with beam FWHM 30 arcmin (dash-dot), but each added with different noise level (dotted): $N_\l=10^{-1}$ and $10^{-3} (\uks)$. Dash lines are the expected residual from XPS according to Eq.(\ref{result}). Bottom panel shows binned ($\Delta \l=10$) power (plus sign) from average of 6 XPS of input WMAP best-fit $\lcdm$ convolved with W channel FWHM 12 arcmin (dash-dot), and added WMAP simulated noise (dotted).}
\label{simu}
\end{figure}

In Figure \ref{simu} we show the XPS (dots) from two simulated maps. Top panel shows Gaussian random signal with power spectrum $\cl=10^{4} \l^{-2}(\uks)$ convolved with beam FWHM 30 arcmin in both maps, the theoretical power spectrum of which is shown in dash-dot curve. They are added with pixel (white) noise signals with differnt levels $N_\l=10^{-1}$ and $N'_\l=10^{-3} (\uks)$. The residual from XPS (i.e. Eq.(\ref{result})) is plotted in dash line. Note that the residual in employing XPS on maps with different noise levels is proportional to $\sqrt{N_\l N'_\l}$, which reflects the real situation of  crossing between different frequency bands such as WMAP V and W band. In bottom panel we simulate 4 WMAP W channel DA maps, convolving with beam FWHM 12 arcmin before adding WMAP simulated noise. There are 6 XPS and the binned ($\Delta \l=10$) retrieved power spectrum is denoted with plus sign. One can see the residual from the noise is the main error from XPS, even if it's not white noise.

\section{Application of cross power spectrum}
One of the applications of XPS is to retrieve window function from bright extragalactic point sources. The standard way of measuring the window function in CMB experiments is via measuring planets such as Jupiter \citep{wmap1ybeam,wmap5ybeam}. 

The importance of window function can be illustrated in the following simple example. The final step of CMB power spectrum estimate often involves deconvolution as the retrieved singal (either via ILC method or foreground template fitting) is also convolved with a common window function. The final power spectrum, after deconvolution, can be written as $\dtq= (2\pi)^{-1} \l (\l+1) \cl^{\rm sm} \exp[b^2 \l (\l+1)]$, where $\cl^{\rm sm}$ is the (retrieved) CMB power spectrum and $b = (\sqrt{8 \ln 2})^{-1}{\rm FWHM}$. An error in deconvolution scale $\Delta b$ produces an error in the power $\Delta(\dtq) = 2 b \, \Delta b \, \l(\l+1) \dtq$, so, for example, deconvolution with $1^\circ$ FWHM on a smoothed map by $59'$ (i.e. overestimated by a mere 1 arcmin) shall overestimate the power at first Doppler peak by $\Delta(\delta T_{220}^2)/ \delta T_{220}^2\simeq 8.9\%$ and the error goes with $\sim \l^2$ for higher $\l$. Although in practice deconvolution involves estimating the inverse covariance matrix, it  demonstrate that the window function is one of the most crucial issue in CMB power spectrum estimation. The sensitivity of window function is also investigated in \citet{shanks}.

Based on flat-sky approximation and Fourier transform, we can estimate the window function from bright point sources by taking two square patches of sky where there is a bright point source in the center:   $T_1({\bi k})= \alpha b({\bi k}) +c({\bi k})+n({\bi k})$, $T_2({\bi k})= \beta b({\bi k}) +c'({\bi k})+n'({\bi k})$, where ${\bi k}$ is the Fourier wavenumber, $c$, $c'$ and $n$, $n'$ represent different CMB($+$foreground) and noise, respectively. The CMB and noise are not correlated in different parts of the sky so they are now the ``junk'' in XPS whereas the bright point sources each in the center of the patches is manifestation of the inflight beam profile and thus are correlated signal. The window function is then $W_k=\sum b^2({\bi k})$ with a rescaling relation $\l=2\pi k/L$, where $L$ is the size of the patch.

\begin{figure}
\plottwo{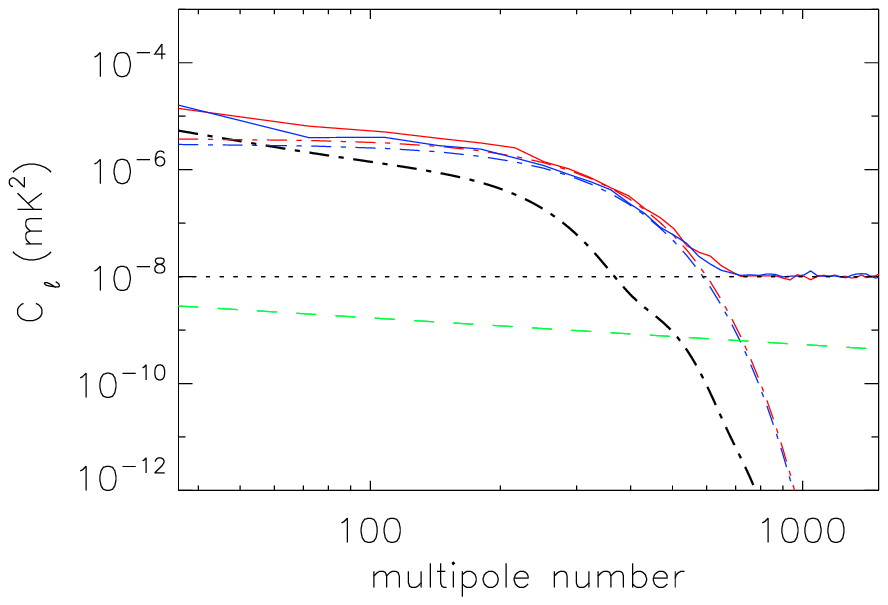}{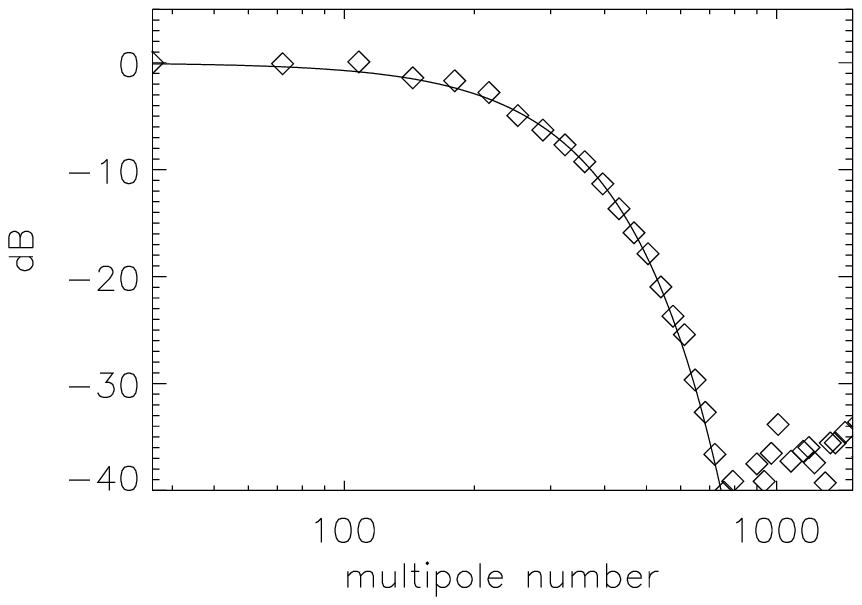}
\caption{Retrieving window function from bright point sources via XPS. Top panel shows the power spectra (solid red and blue curves) of 2 $10^\circ \times 10^\circ$ patches taken from a simulated full-sky map of Planck 30 GHz channel. In both patches CMB and a bright point source located at the center are convolved with 33 arcmin FWHM. Dash-dot black curve denotes the theoretical power spectrum of beam-convolved CMB from best-fit $\Lambda$CDM model whereas dash-dot red and blue the point sources with amplitude 47$\sigma$ and 40$\sigma$, respectively, where $\sigma^2$ is the variance of the CMB of the patch. Dotted line is the noise level $10^{-2} \uks$ and  dash green line is the expected residual of the XPS from noise. In bottom panel we show the input window fucntion (solid curve) in dB and the retrieved window function (diamond sign), which is better than $-35$dB as indicated in green line on top panel.}
\label{beam}
\end{figure}

We simulate full-sky CMB signal with WMAP best-fit $\Lambda$CDM model on Planck 30 GHz channel and add 2 bright point sources on different parts of the sky, which is then convolved with beam 33 arcmin FWHM before adding noise with $N_\l=10^{-2} \uks$. We extract $10^\circ \times 10^\circ$ with the point source at the center. The point sources have amplitudes 40 and $46 \sigma$, where $\sigma^2$ is the variance of the CMB. In Fig.\ref{beam} we present the simulation and the retrieving capability of the window function. One can see that XPS extended the estimate more than 10 dB.

\section{Window Functions of the WMAP Difference Assemly and Frequency Band Maps}
Having demonstrated that the window function can be retrieved with XPS from simulated bright point sources, we now employ the same method on WMAP DA maps: K1, Ka1, Q1, Q2, V1, V2, W1--W4. For each DA map, we extract the square patches where the top 2 brightest point sources are in the center of the patche. 

Care has to be taken when one chooses the patch size. Fourier transform spreads the variance of the map into the power of different wavelengths. For stationary noise, choosing different patch area : changing from $L_0^2$ to $L^2$ (with the same pixel size) shall shift the level of white spectrum by $(L_0/L)^2$ as the variance is fixed. For a point source, on the other hand, changing the patch size has an {\it extra} effect on the variance by $(L_0/L)^2$, resulting in the large-scale power level of a point source being shifted by $(L_0/L)^4$. Hence, choosing a smaller patch ($L < L_0$) enhances the relative power level of the point source to noise by $(L_0/L)^2$, rendering a better profile of the window function. 

One cannot, however, choose as small sizes as possible. The first multipole number to display the profile of the window function is decided by the size of the patch $L$ via $\l_{\rm first}=2\pi/L$, so one loses the large-scale profile if the patch size is too small. Furthermore, a fixed pixel size sets the maximum multipole number regardless of the patch size. So a smaller patch with a fixed pixel size has fewer Fourier modes with spacing $\Delta \l=2\pi/L$.

\begin{figure}
\epsfig{file=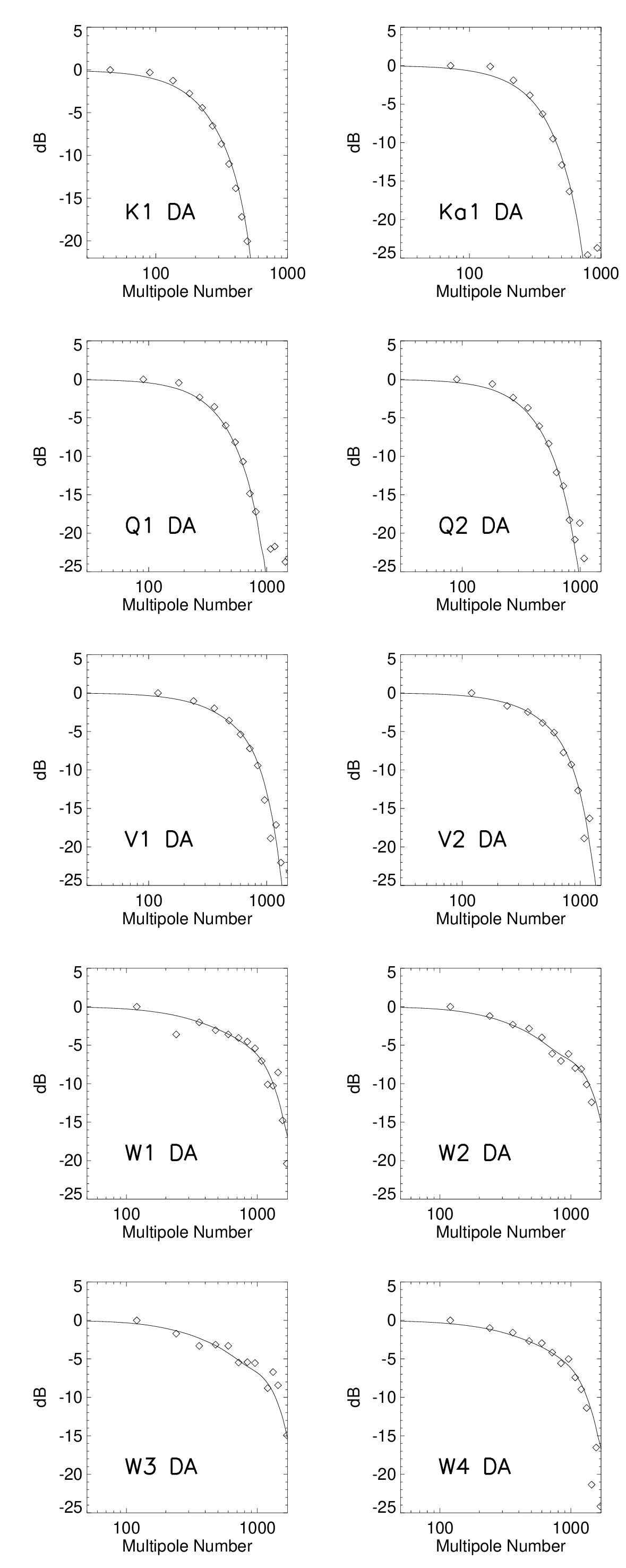,width=8cm}
\caption{The retrieved window functions of the 10 WMAP DA maps. Solid curves are the WMAP window functions for the DA maps and the diamond sign denotes the retrieved ones after crossing two patches of brightest point sources. One can see that the method is doing well on K, Ka, Q and V band DA maps, except for the W band DA maps, which is due to the small beam size and high level noise.}
\label{da}
\end{figure}

We take the patch size $L=8$, 5, 4, 3 and 3 degrees for K, Ka, Q, V and W band respectively and show the XPS results of all 10 DA maps in Fig 6: solid curves are the window functions from WMAP DA maps and diamond sign denotes the retrieved window functions from XPS. One can see in Fig.\ref{da} that the XPS retrieves the window functions nicely, except for the W band 4 DA maps. The reason is that W band not only has the smallest beam size, but also higher noise level than the other bands. 

According to the previous XPS DAs results, our method for retrieving window function from bright point sources is demonstrated useful. We can now retrieve the window functions of the WMAP frequency band maps. The WMAP frequency band maps, together with the corresponding window functions, can be as useful as the DA maps in extracting the CMB power spectrum. The WMAP K, Ka, Q, V and W band have 1, 1, 2, 2 and 4 DA maps respectively, and the frequency band maps are produced from combining the DA maps, thus there is no direct measurement or estimation of the corresponding window functions for each map. One should note that the window functions of the DA maps at the same frequency band do not necessarily have the same profile, which is particularly true for W band (e.g. see W1 and W2 DA window functions in Fig.\ref{da}).

\begin{figure}
\epsfig{file=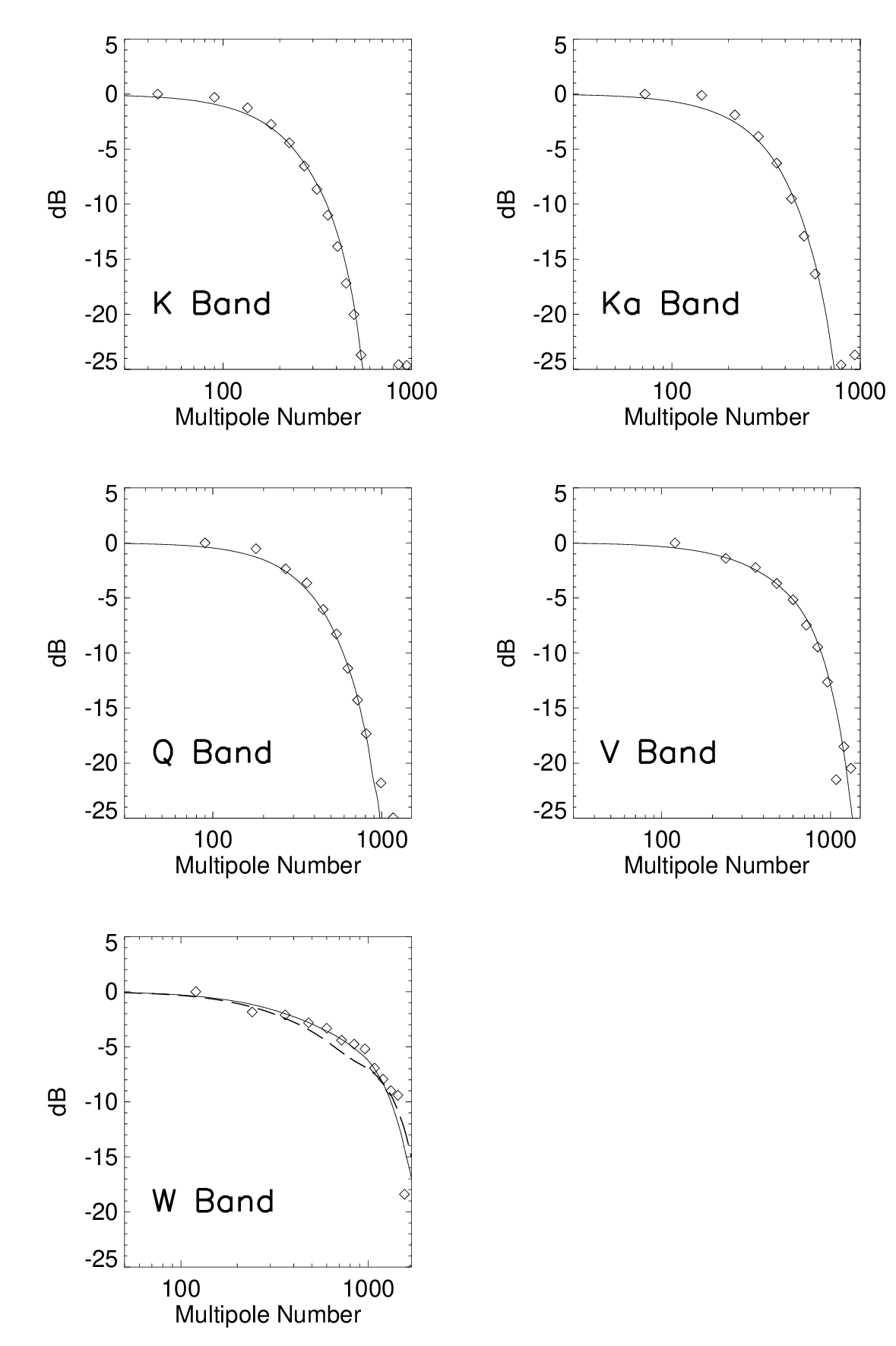,width=9.2cm}
\caption{The retrieved window functions of the WMAP frequency band map. We employ the same procedure as that in retrieving the DA maps in Fig.\ref{da}. The diamond sign denotes the retrieved window function and, for comparison, we put the window function profiles of the DA for comparison. W band is compared with W1 (solid curve) and W2 (long dashe) window function.  }\label{frequencyband}
\end{figure}

In order to retrieve the window functions of the frequency band maps, we perform the same procedure as that on the DA maps in the previous section by choosing the same patches of the sky and patch size for the corresponding band. The retrieved window functions are showed in Fig.\ref{frequencyband}, and for comparison, we put the window function profiles of the DA maps for the corresponding band. Though there are 2 DA maps for Q and V band, the profiles are close to each other, and therefore one can see our retrieved window function for the frequency band maps are close to that of DA maps'. For W band, however, the difference in DA maps is pronounced in that the retrieved W band window function is compared with that of W1 DA map (solid curve) and W2 DA map (long dash). The profile of W3 and W4 are not plotted as they are close to that of W2 and W1 respectively. One can see that the window function profile of the W frequency band map is close to that of W1 (and W4), not W2 (and W3).

\section{Conclusion and Discussion}
In this paper we discuss the limit of the widely-used XPS and we demonstrate that it has the ability to decrease uncorrelated ``junk'' signal by a factor inversely proportional to the square root of the number of modes involved in the summation. It indeed can decrease the uncorrelated signal on small scales in the power spectrum, but has limited ability in eliminiating that of large scales. We then employ XPS to retrieve window functions via extragalactic point sources. One of the crucial steps for estimating CMB power spectrum is to know about the window function that convolves the observed signal. In this case the ``junk'' signal is composed of CMB and pixel noise. We take WMAP DAs as test samples first and employ XPS on extragalactic bright point sources to extract window function. We demonstrate it a useful method to extract window function without measuring planets for DAs. For frequency band maps  combined from DA maps, we apply the same method to extract the window functions, which are unknown prior to our investigation. Our method will be useful for the upcoming Planck data.

\acknowledgments
 We acknowledge the use of \healpix\footnote{{\tt http://healpix.jpl.nasa.gov/}} package \citep{healpix}  and the use of \glesp\footnote{{\tt http://www.glesp.nbi.dk/}} package. The author would like to thank Peter Coles and Dipak Munshi for useful discussions.



\clearpage


\begin{thebibliography}{}
\bibitem[Bennett et al. (2003)]{wmap1yfg} Bennett, C.L. et al., 2003, ApJS, 148, 97 
\bibitem[Chiang, Naselsky and Coles (2009)]{cc} Chiang, L.-Y., Naselsky, P.D., Coles, P., 2009, ApJ, 694, 339 
\bibitem[Doroshkevich et al. (2005)]{glesp} Doroshkevich, A.G. et al., 2005, IJMPD, 14, 275  
\bibitem[Gorski et al. (2005)]{healpix} Gorski, K. M., Hivon, E., Banday, A. J.,Wandelt, B. D., Hansen, F. K., Reinecke, M., \& Bartelmann, M., 2005, ApJ, 622, 759 
\bibitem[Hill et al. (2009)]{wmap5ybeam} Hill, R. et al., 2009, ApJS, 180, 246
\bibitem[Hinshaw et al. (2003)]{wmap1ypower} Hinshaw, G. et al., 2003, ApJS, 148, 135 
\bibitem[Hinshaw et al. (2007)]{wmap3ytem} Hinshaw, G. et al., 2007, ApJS, 170, 288 
\bibitem[Hughes (1995)]{hughes}Hughes, B. D., 1995, Random Walks and Random Environments: Volume 1: Random Walks, Oxford University Press 
\bibitem[Naselsky et al. (2005)]{naselsky}Naselsky, P.D., Chiang L.-Y., Olesen, P., Novikov, I., 2005, PRD, 72, 063512 
\bibitem[Page et al. (2003)]{wmap1ybeam} Page, L. et al., 2003, ApJS, 148, 39 
\bibitem[Pearson (1905)]{pearson} Pearson, K., 1905, Nature, 72, 294
\bibitem[Larson et al. (2010)]{wmap7ypower} Larson, D. et al., 2010, ApJS submitted 
\bibitem[Nolta et al. (2009)]{wmap5ypower} Nolta, M. et al., 2009, ApJS, 180, 296
\bibitem[Rayleigh (1905)]{rayleigh} Rayleigh, J.S., 1905, Nature, 72, 318 
\bibitem[Stannard \& Coles (2005)]{coles}Stannard, A., Coles, P., 2005, MNRAS, 364, 929 
\bibitem[Sawangwit \& Shanks (2010)]{shanks} Sawangwit, U., Shanks, T., 2010, MNRAS accepted
\bibitem[Springer (1979)]{springer}Springer, M.D., 1979, The Algebra of Random Variables, John Wiley \& Sons
\bibitem[White, Krauss and Silk (1993)]{cv} White, M., Krauss, L.M., Silk, J., 1993, ApJ, 418, 535 
\end{thebibliography}
\end{document}